\documentclass[preprint,aps]{revtex4}
\usepackage{color}
\usepackage{amsmath,amssymb,amsfonts,dcolumn,color,graphicx,graphics,latexsym,epsfig}
\def\beq{\begin{equation}}
\def\eeq{\end{equation}}
\def\barr{\begin{array}}
\def\earr{\end{array}}

\begin{document}

\title{Can extra dimensional effects allow wormholes without exotic matter?}

\author{Sayan Kar${}^\#$, Sayantani Lahiri${}^{\dagger,\ddagger}$ and
Soumitra SenGupta ${}^{*}$}
\email{sayan@iitkgp.ac.in,sayantani.lahiri@gmail.com,tpssg@iacs.res.in}
\affiliation{${}^\#$Department of Physics {\it and} Center for Theoretical Studies \\Indian Institute of Technology, Kharagpur, 721 302, India,}
\affiliation{${}^{\dagger}$Institute for Physics, University Oldenburg, D-26111 Oldenburg, Germany,}
\affiliation{${}^{\ddagger}$ZARM, University of Bremen, Am Fallturm, 28359 Bremen, Germany}
\affiliation{${}^{**}$Department of Theoretical Physics, Indian Association for the
Cultivation of Science
\\ 2A and 2B Raja S.C. Mallick Road, Jadavpur, Kolkata 700 032, India.}

\begin{abstract}
\noindent  
We explore the existence of Lorentzian wormholes 
in the context of an effective on-brane, scalar-tensor theory of gravity. 
In such theories, the timelike convergence condition, which is
always violated for wormholes, has contributions, via the field equations,
from on-brane matter as well as from an effective geometric stress energy 
generated by a bulk-induced radion field.
It is shown that, for a class of wormholes, the required on-brane 
matter, as seen by an on-brane observer in the Jordan frame, is not exotic 
and does not violate the
Weak Energy Condition. The presence of the 
effective geometric stress energy in addition to on-brane matter, 
is largely responsible for 
creating this intriguing possibility.   
Thus, if such wormholes are ever found to exist in the 
Universe, they would clearly provide pointers towards the existence of a
warped extra dimension as proposed in the two-brane model of Randall and 
Sundrum.  
\end{abstract}


\maketitle

\newpage

\noindent {\em Introduction:} Ever since Einstein and Rosen \cite{einstein} proposed the Einstein-Rosen bridge
and Wheeler \cite{wheeler}  coined the term {\em wormhole}, 
such geometries have 
been of great interest both in physics as well as in science fiction.
One of the first non-singular wormhole solutions using a wrong-sign scalar field
was found by Ellis in 1973 \cite{ellis}. 
Subsequently,
in the late 1980s, Morris, Thorne and Yurtsever \cite{mty} 
came up with a time-machine
model using wormholes. This led to a host of articles on wormholery \cite{wormholes}.
However, a major drawback of all work on static wormholes has been
the proven fact that wormholes must violate the energy conditions \cite{mt}.
Energy Conditions \cite{hawk,wald} 
are known to be
sacred because they point towards physical requirements on matter.
For example, we know that if the Weak Energy Condition is violated it
implies a negative energy density in some frame of reference \cite{mt}. 
Within the framework of General Relativity, it is impossible to
construct a static wormhole without violating the energy conditions.
Thus, classically, such spacetimes cannot exist. 

\noindent Many arguments may be given in support of the violation of
the energy conditions. Some of them invoke quantum fields in curved spacetime
--the renormalised stress energy tensor is known to have such violations \cite{qft} .
Attempts have been made towards avoiding violations or justifying them,
in time-dependent spacetimes \cite{sk}and in alternative theories of gravity
\cite{alt}. Further, there have been proposals regarding restricting
the violation of the energy conditions over arbitrarily small regions \cite{vkd}.

\noindent In our work here we try to address the issue in a different
way. We ask whether the existence of extra dimensions can, in some way
lead to the existence of Lorentzian wormholes which do not violate any
energy condition. In other words, using the warped braneworld picture \cite{rs}, can we
say that an observer sitting on a 3-brane does not see a violation of
the energy conditions for matter that threads a possible wormhole
geometry? To make things more concrete, we use the effective, on-brane
scalar -tensor theory constructed by Kanno and Soda \cite{ks} in the context 
of the
two-brane Randall-Sundrum model, in a higher dimensional bulk spacetime \cite{rs}.
In such a theory, the scalar radion field which encodes information about
the extra dimensions and branes, plays a crucial role. The radion
is a measure of the proper distance between the branes. We find that
for a class of wormhole geometries, the radion is everywhere finite and
non-zero and the on-brane matter threading the wormhole is perfectly
normal without any violation of the energy conditions. Therefore, if
we ever see such a wormhole, we may be able to prop it up as a support
for the existence of warped extra dimensions as proposed in the
 two-brane Randall-Sundrum model.
We now elaborate in detail on this exciting possibility.

\

\noindent {\em Low energy, effective, on-brane gravity:} Let us begin by writing down the four-dimensional, 
effective scalar-tensor theory due to Kanno-Soda \cite{ks}. 
The higher dimensional bulk spacetime is five dimensional with a
warped extra dimension and two 3-branes located at $y=0$ and $y=l$, where
$y$ denotes the extra dimension. The effective theory is a valid low energy
theory as long as the on-brane matter energy density is much less than
the brane tension. It is therefore clear that near singularities the
theory will break down. However, since we are dealing with non-singular 
solutions (finite energy density and pressures) in
this article the effective theory is valid everywhere. 
The field equations for this effective theory are given as:
\begin{eqnarray}
G_{\mu \nu} = \frac{{\bar\kappa}^2}{l\Phi} T^{b}_{\mu \nu} + \frac{{\bar\kappa}^2\,(1 + \Phi)}{l\Phi} 
T^{a}_{\mu \nu} + \frac{1}{\Phi}\left ({\nabla}_{\mu} {\nabla}_{\nu} \Phi - g_{\mu \nu} {\nabla}^{\alpha} {\nabla}_{\alpha} \Phi\right ) \nonumber 
\\  -\frac{3}{2\Phi(1+\Phi)} \left ({\nabla}_{\mu}\Phi{\nabla}_{\nu}\Phi
-\frac{1}{2} g_{\mu \nu} {\nabla}^{\alpha}\Phi{\nabla}_{\alpha} \Phi\right )  \label{EE}
\end{eqnarray}
Here $g_{\mu\nu}$ is the on-brane metric and the covariant
differentiation is defined with respect to  $g_{\mu\nu}$. 
${\bar\kappa}^2$ is the $5D$ gravitational coupling constant.
$T^{a}_{\mu \nu}$, $T^{b}_{\mu \nu}$ are the
stress-energy on the Planck brane and the visible brane respectively.
The appearance of $T^{a}_{\mu\nu}$ (matter energy momentum on the `a' brane)
in the field equations on the `b' brane, inspired the usage of the term
`quasi-scalar-tensor theory'in this context\cite{ks}. Assuming
$T^a_{\mu\nu}=0$ we
have a usual scalar-tensor theory. We will assume no matter on the
`a' brane in our future discussion. 

\noindent Note the scalar $\Phi$ which appears in the equations.
$\Phi$ is known as the radion field and it is a measure of the
distance between the two branes. $\Phi$ depends on the brane coordinates
which are collectively referred as {\sf x}. We have 
\begin{equation}
\Phi ({\sf x}) = e^{2\frac{d({\sf x})}{l}}-1
\end{equation}
where $d({\sf x})$ is the proper distance between the two branes
given as
\begin{equation}
d({\sf x})\,=\, \int^{l}_{0}e^{\phi ({\sf x})} dy
\end{equation}
with $\phi$ appearing in the five dimensional line element
\begin{eqnarray}
ds_5^2= e^{2\phi({\sf x})} dy^2 + {\tilde g}_{\mu\nu} (y, x^{\mu}) dx^\mu dx^\nu
\end{eqnarray}
The scalar radion satisfies the field equation
\begin{equation}
{\nabla}^{\alpha}{\nabla}_{\alpha}\Phi =\frac{{\bar\kappa}^2}{l}\frac{T^a + T^b}{2\omega+3} - 
\frac{1}{2\omega +3} \frac{d\omega}{d\Phi} ({\nabla}^{\alpha}\Phi)(
{\nabla}_{\alpha}\Phi )
\end{equation}
with $T^{a}$, $T^{b}$ being the traces of energy momentum tensors on
Planck (`a') and visible (`b') branes, respectively.
The coupling function $\omega({\Phi})$ is expressed in terms of $\Phi$ as
\begin{equation}
\omega (\Phi) = -\frac{3\Phi}{2(1+\Phi)}
\end{equation}
It is crucial to have the following physical 
conditions on the $\Phi ({\sf x})$ (or the $d({\sf x})$).

\noindent $\bullet$ $\Phi$ is never zero.

\noindent $\bullet$ $\Phi$ does not diverge to infinity at any finite value of the brane coordinates.

\noindent The above conditions imply that the branes do not collide and 
nowhere does the brane separation become infinitely large. One may
say that if the above conditions are obeyed by $\Phi$ we have a stable
radion.

\

\noindent {\em Energy conditions:} It is easy to see that the field equations
(Eqn. (1)) for the line element
can be formally written as:
\begin{eqnarray}
G_{\mu \nu} = \frac{{\bar\kappa}^2}{l\Phi} T^{b}_{\mu \nu}  
+ \frac{1}{\Phi} T^{\Phi}_{\mu\nu}
\end{eqnarray}
where $T^{\Phi}_{\mu\nu}$ constitutes the third and fourth terms 
(without the $\frac{1}{\Phi}$ factor) in the R. H. S. of Eqn. (1).
Recall that the Raychaudhuri equation for the expansion $\Theta$ of 
timelike geodesic congruences is given as:
\begin{equation}
\frac{d\Theta}{d\lambda} + \frac{1}{3}\Theta^2 +\Sigma^2 -\Omega^2 = -R_{\mu\nu}
u^\mu u^\nu
\end{equation}
where $u^\mu$ is the tangent vector to the central geodesic in the
congruence, $\Sigma^2= \Sigma_{ij}\Sigma^{ij}$ ($\Sigma_{ij}$ is the shear),
$\Omega^2 = \Omega_{ij}\Omega^{ij}$ ($\Omega_{ij}$ is the rotation) 
and $\lambda$ is the affine parameter.
We know \cite{hawk,wald} that geodesics focus within a finite value of the
affine parameter provided $R_{\mu\nu}u^\mu u^\nu\ge 0$ (the timelike convergence condition). 
In General Relativity, using the Einstein field equations, the timelike
convergence condition becomes the Strong Energy Condition
$(T_{\mu\nu}-\frac{1}{2}g_{\mu\nu}T)u^\mu u^\nu\ge 0$. Other versions of
the energy conditions include the Weak Energy Condition 
$T_{\mu\nu}u^\mu u^\nu\ge 0$ or the Null Energy Condition, 
$T_{\mu\nu} k^\mu k^\nu \ge 0$ ($k^\mu$ being the tangent to null
geodesics).
Such energy conditions are deemed important 
since they lead to physical requirements on matter.
For example, the Weak Energy Condition for a diagonal energy momentum tensor
reduces to the set of inequalities $\rho\geq 0,\rho+\tau\geq 0, \rho+p\geq 0$
where $\rho$, $\tau$ and $p$ correspond to the energy density and the
radial and tangential pressures, respectively. It can be shown that
this set of WEC inequalities imply that the energy density is never
negative in any frame of reference \cite{mt}. Thus, independent of
the timelike convergence condition, we can assume these conditions as
requirements that all known energy momentum tensors of matter must
obey. 

\noindent In a theory of gravity which is not General Relativity,
the relation between the timelike convergence condition and the
energy condition (say WEC) is not direct \cite{scapo}. Let us now look at this
aspect from the standpoint of the effective scalar-tensor theory we are 
considering. We work in the Jordan frame. We also assume that the line 
elements we will be
considering are those for which the Ricci scalar $R=0$. The convergence condition then becomes
\begin{equation}
R_{\mu\nu}u^\mu u^\nu = \frac{{\bar\kappa}^2}{l\Phi} T^{b}_{\mu \nu} u^\mu u^\nu+ \frac{1}{\Phi} T^{\Phi}_{\mu\nu} u^\mu u^\nu \ge 0
\end{equation}
It therefore becomes possible to satisfy the convergence condition 
but, at the same time, have a violation of the WEC for $T^b_{\mu\nu}$.
Similarly one can violate the convergence condition but still satisfy
the WEC. Such freedom arises entirely  due to the presence  
of the extra term, i.e. the
effective geometric stress energy, $T^{\Phi}_{\mu\nu}$, due to the 
radion scalar. 

\noindent The important question here is, what does an
observer on the brane see? Obviously, such an observer in the Jordan frame,
will only
see $T^b_{\mu\nu}$ \cite{scapo}. But the focusing, defocusing of geodesic
congruences will be decided by the nature of $R_{\mu\nu} u^\mu u^\nu$.
Why isn't the radion effective stress energy visible and measurable 
to the brane observer? The answer is similar to the motivation behind
introducing a scalar field, in the original Brans-Dicke theory, where
it was responsible for generating the gravitational constant $G$ \cite{weinberg}. Here too, the
presence of the radion signals the existence of extra dimensions and has 
nothing to do with the ordinary matter which is seen by the Jordan frame
observer.  

\noindent We will exploit the above arguments while constructing our on-brane 
Lorentzian wormhole spacetime.

\

\noindent {\em The Lorentzian wormhole:} It is known that, in General Relativity,
 static wormholes cannot satisfy the
energy conditions on matter. The wormhole throat acts as a defocusing lens
which leads to the violation of the timelike/null convergence condition.
We shall consider here a known wormhole solution with $R=0$ \cite{dkmv}.
In Schwarzschild coordinates, such a wormhole is given by the
line element
\begin{equation}
ds^2 = -\left (\kappa +\lambda \sqrt{1-\frac{2m}{r'}}\right )^2 dt^2 +
\frac{dr'^2}{1-\frac{2m}{r'}} + r'^2 \left (d\theta^2 +\sin^2\theta d\phi^2\right )
\end{equation}
Note there is no horizon or singularity
in this line element (i.e. $g_{00}$ never equal to zero) as long as $\kappa$,
$\lambda$ are both either positive (or negative) with $\vert \kappa\vert >
\vert \lambda \vert$. In our work here, we choose  
$\kappa>\lambda>0$. The spatial section 
of the geometry is identical to that of Schwarzschild spacetime.
$r'=2m$ is the location of the wormhole throat. 
Using the isotropic coordinate $r$ where 
$r' = r\left(1+\frac{m}{2r}\right )^2$, the line element becomes
\begin{equation}
ds^2 = -\left (\kappa + \lambda \frac{1-\frac{m}{2r}}{1+\frac{m}{2r}}\right )^2 dt^2 + \left (1+\frac{m}{2r}\right )^4 \left (dr^2 + r^2 d\theta^2 + r^2\sin^2 \theta d\phi^2 \right )
\end{equation}

\noindent The fact that the Ricci scalar is identically zero inspires us to
see if this is a viable line element in the Kanno-Soda effective theory of 
gravity. The general form of a spherically symmetric static line element in
isotropic coordinates is assumed as,
\begin{equation}
ds^2=-\frac{f^2(r)}{U^2(r)} dt^2 +U^2(r) \left [ dr^2 + r^2 d\theta^2
+r^2\sin^2\theta d\phi^2 \right ]
\end{equation}
where $U(r)$ and $f(r)$ are the unknown functions to be determined by solving
the field equations.
Using the above line element ansatz and the assumption that
$\Phi$ is a function of $r$ alone, we obtain the following field equations,
\begin{eqnarray}
-2 \frac{U''}{U} +\left (\frac{U'}{U}\right )^2 - 4\frac{U'}{Ur} =
-\frac{\Phi'^2}{4\Phi(1+\Phi)} + \left (\frac{U'}{U} -\frac{f'}{f}\right ) 
\frac{\Phi'}{\Phi} +\frac{{\bar\kappa}^2}{l \Phi} \rho  \\
-\left (\frac{U'}{U}\right )^2 + 2\frac{f'}{f} \left (\frac{U'}{U} + \frac{1}{r}\right ) =  
-\frac{3\Phi'^2}{4\Phi(1+\Phi)} - \frac{U'}{U}\frac{\Phi'}{\Phi} -
\frac{2\Phi'}{\Phi r} - \frac{f'}{f} \frac{\Phi'}{\Phi} +\frac{{\bar\kappa}^2}{l\Phi} \tau \\
\left (\frac{U'}{U}\right )^2 + \frac{f''}{f} -2\frac{f'}{f} \frac{U'}{U} +\frac{f'}{f}\frac{1}{r}  =  
\frac{\Phi'^2}{4\Phi(1+\Phi)} + \frac{U'}{U}\frac{\Phi'}{\Phi} +
\frac{\Phi'}{\Phi r} +\frac{{\bar\kappa}^2}{l\Phi} p
\end{eqnarray}
where $\rho$, $\tau$ and $p$ correspond to on-brane matter and,
using the tracelessness condition, we have $-\rho+\tau +2 p =0$.
We have absorbed a factor of $U^2$ in the definitions of $\rho$, $\tau$
and $p$.

\noindent On the other hand, the scalar ($\Phi$) field equation becomes
\begin{equation}
\Phi'' + \frac{f'}{f} \Phi' + 2\frac{\Phi'}{r} = \frac{\Phi'^2}{2(1+\Phi)}
\end{equation}
The above equation can be integrated once to get
\begin{equation}
\frac{\Phi'}{\sqrt{1+\Phi}} = \frac{2C_1}{r^2 f}
\end{equation}
where $C_1$ is a positive, non-zero constant. Notice that radion 
field equation has no contribution from on-brane matter, essentially because
$T^b_{\mu\nu}$ is assumed traceless. 

\noindent Further,
the traceless-ness requirement leads to a single equation for the
metric functions, given as
\begin{equation}
\frac{U''}{U} + \frac{f''}{f} -\frac{f'}{f}\frac{U'}{U} + 2\frac{f'}{f r}
+ 2 \frac{U'}{U r} =0
\end{equation}

\noindent For the $R=0$ line element mentioned earlier (see Eqns. (10), (11)), 
and using isotropic coordinates, we have
\begin{equation}
f(r) = \left (1+\frac{m}{2r}\right ) \left [\kappa\left (1+\frac{m}{2r}\right )
+\lambda \left( 1-\frac{m}{2r}\right )\right ]
\end{equation}
\begin{equation}
U(r) = \left (1+\frac{m}{2r}\right )^2
\end{equation}
\noindent We can check that the above $f$ and $U$ satisfy the
traceless-ness condition ($R=0$) given in Eqn. (18). Earlier work
on similar $R=0$ solutions (with $f=1$) in the context of KS effective theory
can be found in \cite{skslssg}.

\noindent To proceed we now need to know the $\Phi(r)$.
With the above $f(r)$ we can obtain $\xi = \sqrt{1+\Phi}$ quite easily.
This leads to
\begin{equation}
\xi' = \frac{C_1}{r^2 (\kappa-\lambda)}\frac{1}{(1+\frac{m}{2r})(q+\frac{m}{2r})}
\end{equation}
and
\begin{equation}
\xi = \frac{C_1}{m\lambda} \ln \frac{2rq+m}{2r+m} + C_4
\end{equation}
where $q=\frac{\kappa+\lambda}{\kappa-\lambda}$. To get a well-behaved radion,
we need a $\Phi$ which is never zero or infinite. 
The above requirement implies that we choose $q>1$ with $m$, $C_4$ and $C_1$ positive and non-zero.
It is useful to note that $\Phi$ can be greater than zero even if
$C_4=0$ as long as $q>1$ and the other constants are suitably
adjusted. We will use this fact later, in this article. 

\noindent If we assume $\kappa=0$ and $\lambda=1$, then we have
Schwarzschild geometry ($q=-1$). In this case, the $\xi(r)$ turns out
to be
\begin{equation}
\xi= \frac{C_1}{m} \ln \frac{2r-m}{2r+m} + C_4
\end{equation}
It is easy to see that there is always a zero (infact two zeros)
of $\Phi(r)$ for Schwarzschild. The only possibility then is to take
a constant $\Phi$, which is trivial \cite{grg}.

\

\noindent {\em Checking the Weak Energy Condition:} We now need to verify the nature of the on-brane 
matter that threads the
wormhole geometry. From the field equations given earlier (Eqns. (13, (14), (15)), it is easy to 
obtain the $\rho$, $\tau$ and $p$ and verify the Weak Energy Condition
inequalities.
We shall now explicitly write down the L. H. S. of these inequalities.

\begin{equation}
\frac{{\bar\kappa}^2}{l}\rho = \frac{16 \beta x^4}{m^2 (1+x)^2(q+x)^2}
\left [ \beta + (q-1)\xi\right ]
\end{equation}

\begin{eqnarray}
\frac{{\bar\kappa}^2}{l}\left (\rho +\tau\right )
= \frac{8 x^3}{m^2 (1+x)^2 (q+x)^2} \left [ 4\beta q \xi + q(1+q)-q(1+q) \xi^2
\right.
\nonumber \\
\left. + x\left (8\beta^2 +(1+q)^2-(1+q)^2 \xi^2\right ) \right.\nonumber \\ \left. 
+ x^2 \left (
(1+q) - (1+q) \xi^2 - 4\beta \xi \right ) \right ] 
\end{eqnarray}

\begin{eqnarray}
\frac{{\bar\kappa}^2}{l} \left (\rho+p\right )
=\frac{4 x^3}{m^2(1+x)^2 (q+x)^2} \left [ -4 q \xi \beta + q(q+1)\left (\xi^2-1
\right )\right . \nonumber \\ \left. 
+ x\left (8(q-1)\xi \beta +(1+q)^2 (\xi^2-1)\right )\right.\nonumber \\
\left. + x^2 \left( 4\xi\beta + (1+q)(\xi^2-1)\right ) \right ]
\end{eqnarray}
\noindent In the above expression, we have used the following 
re-definitions.
\begin{equation}
x=\frac{m}{2r} \hspace{0.2in};\hspace{0.2in}
C_1= \alpha m\hspace{0.2in};\hspace{0.2in}
\beta=\frac{\alpha}{\kappa-\lambda}
\end{equation}
\noindent The wormhole throat is at $r=\frac{m}{2}$ (or $r'=2m$).
Hence, the domain of $x$ is from $x=0$ to $x=1$. One can check that
the above stress energy is traceless.

\noindent The radion field as a function of $x$ is given in terms of
$\xi$ where $\xi$ is written as:
\begin{equation}
\xi (x) = \sqrt{1+\Phi} = \frac{\alpha}{\lambda}\ln \frac{q+x}{1+x} +
C_4
\end{equation} 
\noindent We have $\beta = \frac{\alpha}{\kappa-\lambda}$. 
Defining $\frac{\kappa}{\lambda}=\nu$, we get $q=\frac{\nu+1}{\nu-1}$.
Also $\beta=\frac{\alpha}{\lambda}\frac{1}{\nu-1}$. Since
$\nu= \frac{q+1}{q-1}$, we get $\frac{\alpha}{\lambda}= \frac{2\beta}{q-1}$.
Hence all the inequalities as well as the radion field are now
defined in terms of the parameters $q$, $\beta$ and $C_4$.

\noindent We must now explicitly check the Weak Energy Condition inequalities
$\rho \geq 0$, $\rho+\tau\geq 0$ and$\rho+p\geq 0$. To this end, we
plot the graphs of these quantities for some sample values of
the various parameters. A general proof for all parameters is not
easy. We first note that $\rho$ is always greater than zero, irrespective of
our choice of $\beta$, $q$ or $C_4$ as long as $\beta>0$, $q>1$ and $C_4>0$.
Figure 1 shows the plot for $\rho$ as a function of $x$. 
For the other two inequalities, let us look at the values of the
term inside the square brackets in (25) and (26), for $x=0$. Note that the
values are exactly opposite to each other. The relevant term is
\begin{equation}
-4q\beta\xi (x=0) + q(q+1)(\xi^2(x=0)-1)
\end{equation}
which, at $x=0$, is positive in the $\rho+p$ expression and negative in the 
$\rho+\tau$
expression. Therefore, unless this term is identicaly zero at $x=0$, 
there will be a
violation of either inequality in the vicinity of $x=0$. Note, at $x=0$,
by virtue of the overall factor $x^3$, in the full expression, 
the value will be zero. But the approach to zero will be
from the positive side for one inequality and from the negative side for the
other. 

\noindent What happens if we choose $\beta$ such that this term is
identically zero at $x=0$? From the above equation, we find this
sets up a relation between $\beta^2$ and $q$, given as,
\begin{equation}
\beta^2= \frac{(q^2-1)(q-1)}{4\left [(q+1)(\ln \,q)^2- 2 (q-1)\ln \,q\right ]}
\end{equation}
For $q=3$ we have
\begin{equation}
\beta^2= \frac{1}{(\ln\,3)^2-(\ln \,3)}
\end{equation}
With this choice for $\beta$ we will have a zero value for the
term $-4q\beta\xi + q(q+1)(\xi^2-1)$ at $x=0$.  
Hence, we fix the following values for the parameters:
\begin{equation}
m=1 \hspace{0.2in};\hspace{0.2in}
q=3 \hspace{0.2in};\hspace{0.2in}
\beta^2=\frac{1}{(\ln\,3)^2-(\ln \,3)}
\hspace{0.2in};\hspace{0.2in}
C_4=0
\end{equation}
We have taken the positive square root of $\beta^2$.
With the above choices, we now plot the L. H. S. of the
inequalities as functions of $x$. 
Since we are plotting the L. H. S as functions of $x$
we must remember that infinity is at $x=0$ and the wormhole throat
(minimum $r$) is at $x=1$. The domain $0\leq x\leq 1$ covers the
entire domain $\frac{m}{2}\leq r\leq \infty$.

\begin{figure}[h]
\begin{minipage}{18pc}
\includegraphics[width=18pc]{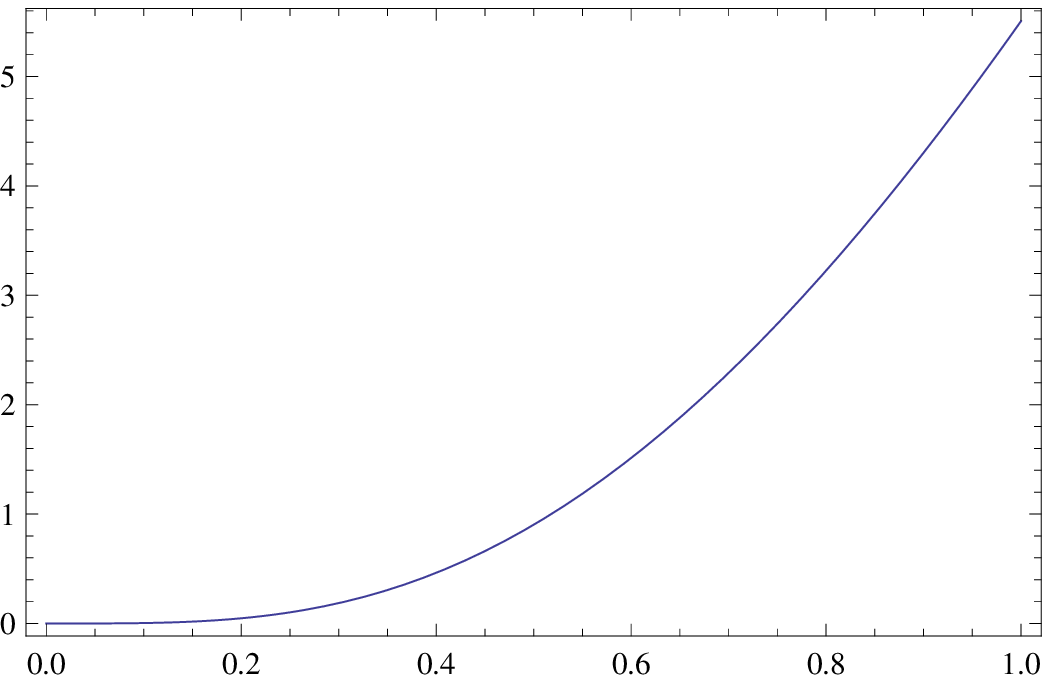}
\caption{$\rho$ vs. $x$}
\end{minipage}
\begin{minipage}{18pc}
\includegraphics[width=18pc]{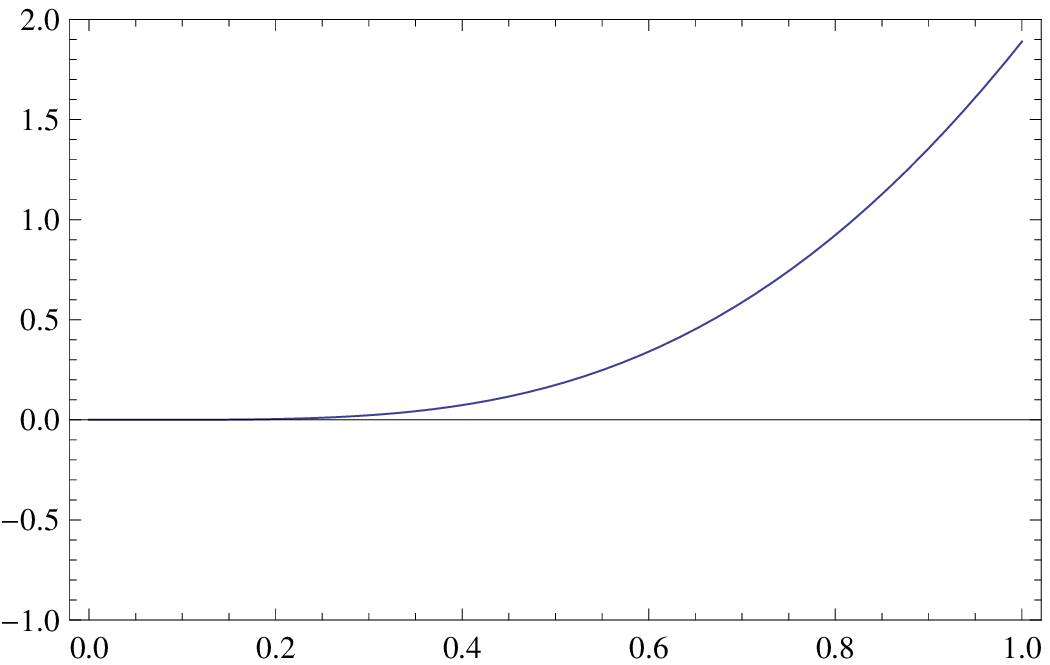}
\caption{$\rho+\tau$ vs. $x$}
\end{minipage}
\end{figure}

\begin{figure}[h]
\begin{minipage}{19pc}
\includegraphics[width=19pc]{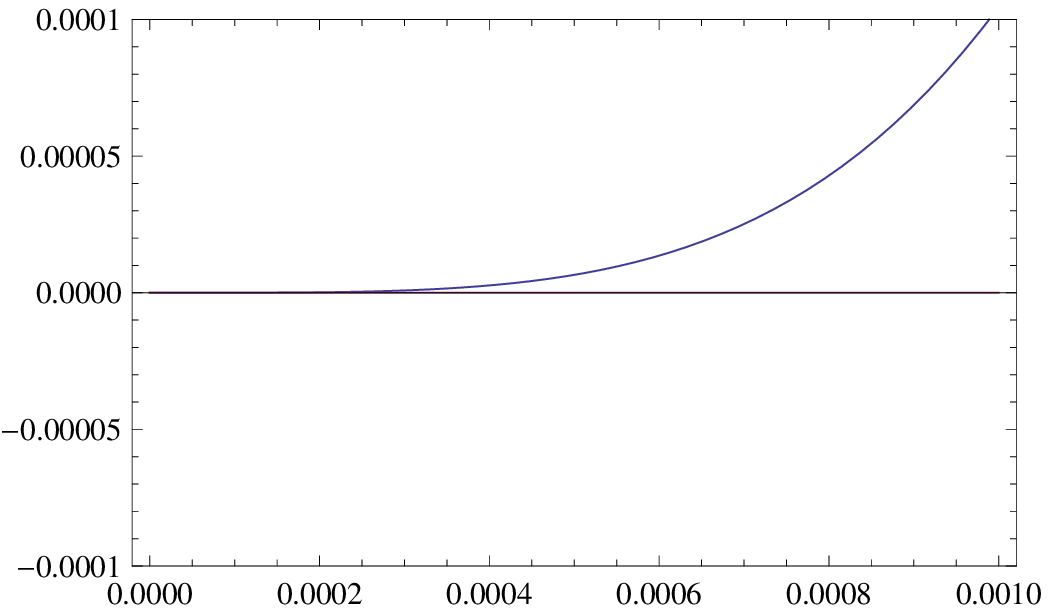}
\caption{${10}^6 (\rho+p)$ vs. $x$}
\end{minipage}
\begin{minipage}{19pc}
\includegraphics[width=19pc]{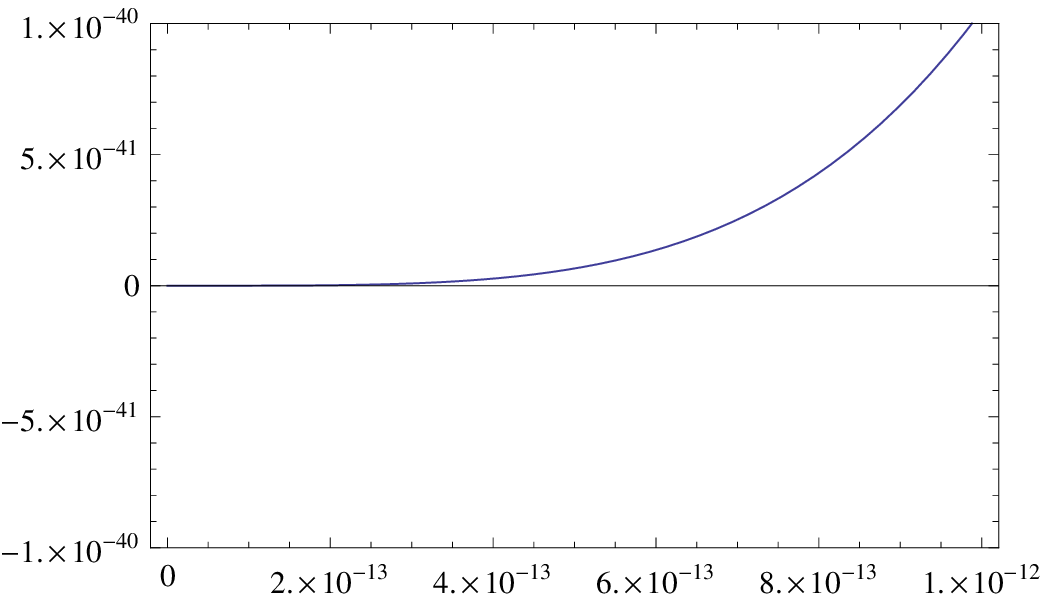}
\caption{${10}^6(\rho+p)$ vs. x, near $x=0$}
\end{minipage}
\end{figure}

\noindent Figure 2 shows a plot of $\rho+\tau$ versus $x$. In Figures 3 and 4, we have plotted $\rho+p$ in various ranges and with overall
constant scale factors, so that we do not miss any negativity. 
The radion field as a function of $x$ is shown in Figure 5. 
The radion is  never zero and it stabilises to an almost constant value for 
large $r$.
The relation between $\beta$ and $q$ is shown in Figure 6.

\begin{figure}[h]
\begin{minipage}{18pc}
\includegraphics[width=18pc]{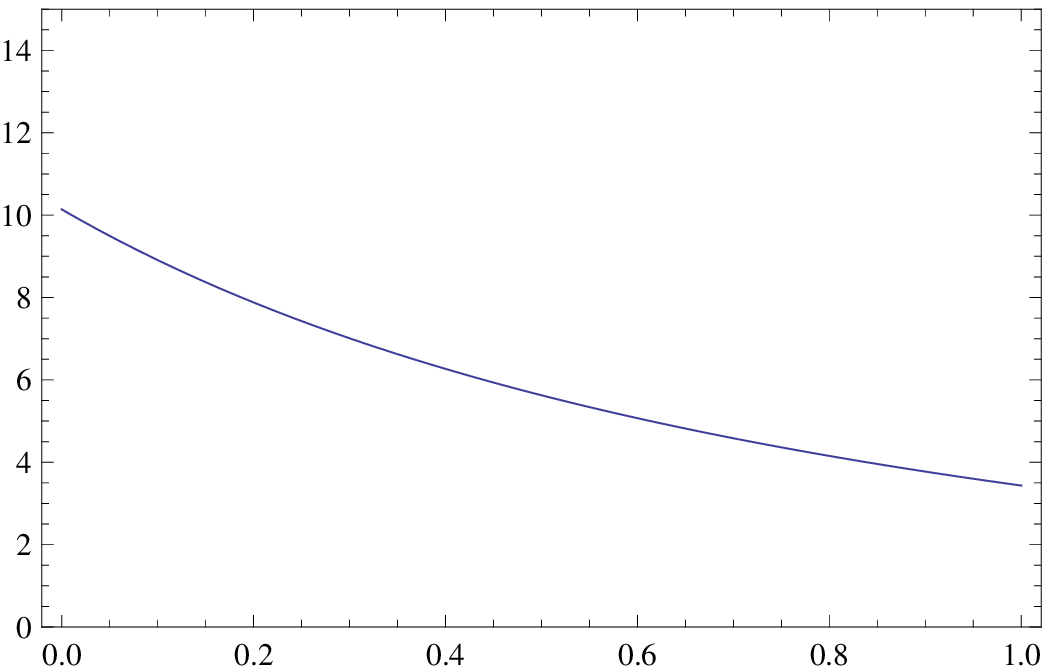}
\caption{Radion $\Phi(x)$ vs. $x$.}
\end{minipage}
\begin{minipage}{18pc}
\includegraphics[width=18pc]{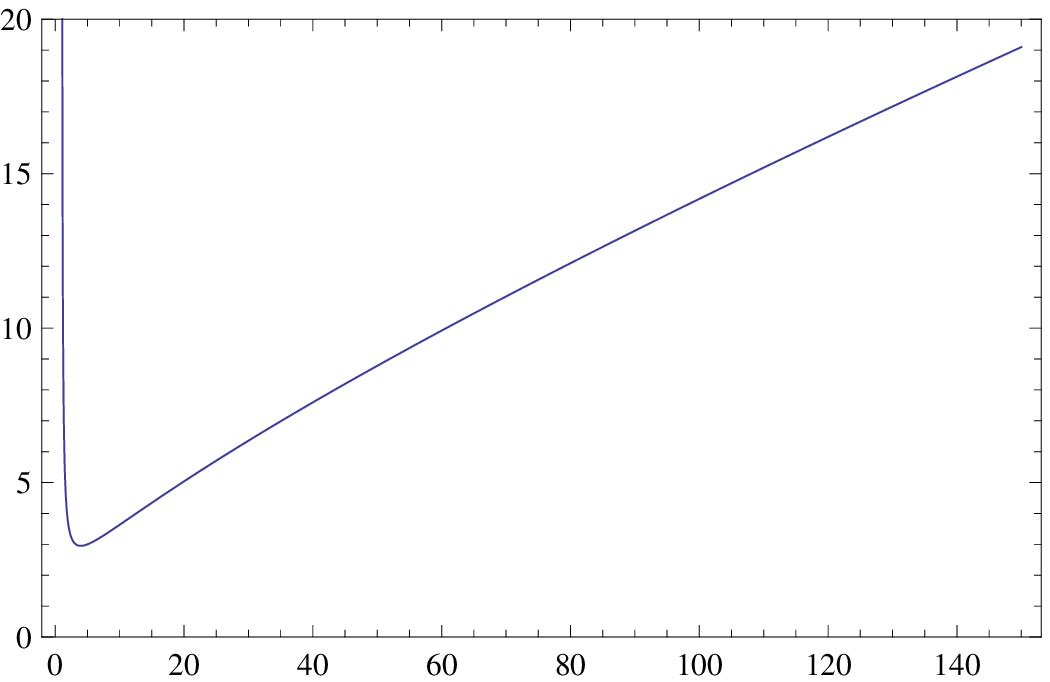}
\caption{$\beta$ vs. $q$.}
\end{minipage}
\end{figure}

\noindent The on-brane matter stress energy is seen to  satisfy
all the energy condition inequalities. This is important because wormholes
are known to violate the energy conditions. It is the effective radion
contribution to the total stress energy which enables the satisfaction of the
Weak Energy Condition for the on-brane matter. In addition, it is also
the requirement of a well-behaved radion which makes this possible.

\noindent One must note that the choice of $\beta$ is very crucial.
Any value of $\beta$ which does not respect the relation between
$\beta$ and $q$ given above will lead to a violation of either the
$\rho+\tau$ inequality or the $\rho+p$ inequality. Further, one must
choose $q>1$ and also $C_4=0$. Fortunately, with $C_4=0$ we still have a
well-behaved radion--the $\Phi$ never equals zero. We mention here
that for $C_4>0$, it is possible to have a well-behaved radion but,
we have checked, using plots over a wide range of values of the various 
parameters, that the energy conditions are indeed violated. At this point,
it seems
unlikely that the choice of $C_4=0$ carries any physical meaning, though
we will not be surprised if it does, in a way of which we are
presently not aware. 

\

\noindent {\em Remarks:} In summary, we  emphasize that we have been able to 
construct an example
of a wormhole which can exist on the visible brane, with matter, as
seen by a Jordan frame, on-brane observer, not violating the Weak Energy
Condition. 
It must be noted that the timelike convergence condition is indeed violated 
--thus, there is no conflict with the geometric conclusions that emerge
from the Raychaudhuri equation for the expansion. 
As stated earlier, the existence
of such a wormhole 
is made possible by the presence of the
radion field which is a measure of the distance between the branes, 
in the two-brane Randall-Sundrum model. 
The radion, being an extra dimensional entity, provides an
effective, geometric stress energy which enables 
the satisfaction of the WEC for on-brane matter. 
In a sense, we have succeeded in `transferring' all the WEC violation
into the radion stress energy, thereby ensuring that the WEC holds for the
observed matter on the brane. 

\noindent Several years ago,
in \cite{bk}, the non-local term in the 
single brane effective theory of Shiromizu-Maeda-Sasaki
\cite{sms} was used to propose $R=0$ wormholes. However, the work presented
here is based on the Kanno-Soda two-brane effective theory which does
not have any non-local term and where, the presence of the extra dimension
is manifest only through the radion scalar. 

\noindent In our investigations here, we have largely focused on a single 
family of wormhole
spacetimes. Obviously, this family is not unique. One can try out
various extensions of this work by looking at other possibilities
with the Ricci scalar $R=0$ and also by removing the $R=0$ restriction.
Further, given the wormhole geometry on the brane it might be
useful to study the behaviour of timelike or null trajectories
and derive interesting physical consequences which can provide
some idea about observable
signatures for such spacetimes.
 
\noindent
Finally, in a broader perspective, the possible existence of a wormhole 
with non-exotic matter could be thought of as
similar to {\em missing energies} in collider phenomenology which are
expected to provide signals of the existence of extra dimensions \cite{lr}.
As noted at the beginning of this article, the existence of such an
on-brane wormhole without exotic matter, can also be a signature of the 
existence of a warped extra dimension.

\

\noindent SL would like to acknowledge the support by DFG Research Training 
Group 1620 `Models of Gravity'.


\begin{thebibliography}{99}
\bibitem{einstein} L. Flamm, Physikalische Zeitscrift {\bf 17}, 448 (1916);  
A. Einstein and N. Rosen, Phys. Rev. {\bf 48}, 73 (1935).
\bibitem{wheeler}  J. A. Wheeler, Phys. Rev., {\bf 97}, 511 (1955).
\bibitem{ellis} H. G. Ellis, J. Math. Phys. {\bf 14}, 104 (1973); Erratum-ibid.
{\bf 15}, 520 (1974); K. A. Bronnikov, Acta Phys. Polon. {\bf B 4}, 251 (1973). 
T. Kodama, Phys. Rev. {\bf D 18}, 3529 (1978);
 H. G. Ellis, Gen. Relativ. Grav. {\bf 10}, 105 (1979). 
\bibitem{mty} M. Morris, K. S. Thorne; U. Yurtsever, Phys. Rev. Letts. {\bf 61},
 1446 (1988). 
\bibitem{wormholes} M. Visser, {\em Lorentzian wormholes: from Einstein to
Hawking}, AIP Press (New York) 1995.
\bibitem{mt} 
M. Morris and K. S. Thorne, Am. J. Phys., {\bf 56}, 395 (1988).
\bibitem{hawk} S. W. Hawking and G. F. R. Ellis, {\em The large scale structure of spacetime}, Cambridge University Press.
\bibitem{wald} R. M. Wald, General Relativity, University of Chicago Press,
First Indian Edition (2006).
\bibitem{qft} H. Epstein, E. Glaser and A. Yaffe, Nuovo Cimento {\bf 36}, 1016 (1965).  
\bibitem{sk} T. Roman, Phys. Rev. {\bf D 47}, 1370 (1993);  
S. Kar, Phys. Rev. {\bf D 49}, 862 (1994); S. Kar and D. Sahdev,
Phys. Rev. {\bf D 53}, 722 (1996); D. Hochberg and M. Visser, Phys. Rev.
Letts. {\bf 81} 746 (1998);  D. Hochberg and M. Visser, Phys. Rev. {\bf D 58}, 
044021 (1998).
\bibitem{alt} D. Hochberg, Phys. Letts {\bf B 251}, 349 (1990); B. Bhawal and
S. Kar, Phys. Rev {\bf D 46}, 2464 (1992); A. G. Agnese and M. LaCamera,
Phys. Rev. {\bf D 51}, 2011 (1995); 
H. Maeda and M. Nozawa, Phys. Rev. {\bf D 78}, 024005 (2008);
P. Kanti, B. Kleihaus, J. Kunz, Phys.Rev.Lett. {\bf 107}, 271101 (2011);
P. Kanti, B. Kleihaus, J. Kunz, Phys.Rev.{\bf  D85}, 044007 (2012);
M. R. Mehdizadeh, M. K. Zangeneh 
and F. S. N. Lobo, Phys. Rev. {\bf D 91}, 084004 (2015); R. Shaikh, arXiv:1505.01314.
\bibitem{vkd}  M. Visser, S. Kar, and N. Dadhich, Phys.
Rev. Lett., {\bf 90}, 201102 (2003). 
\bibitem{rs} L. Randall and R. Sundrum,
Phys. Rev. Lett.  {\bf 83}, 3370 (1999);
{\it ibid} {\bf 83}, 4690 (1999).
\bibitem{ks} S. Kanno, J. Soda, Phys.Rev. {\bf D 66}, 083506 (2002);
T. Shiromizu and K. Koyama, Phys. Rev. {\bf D 67}, 104011 (2003).
\bibitem{scapo} S. Capoziello, F. S. N. Lobo and J. P. Mimoso, 
Phys. Letts. {\bf B 730}, 280 (2014); Phys. Rev. {\bf D 91}, 124019 (2015);
T. Harko, F. S. Lobo, M. Mak, and S. V. Sushkov, Phys.Rev. {\bf D 87},
067504 (2013).
\bibitem{weinberg} S. Weinberg, {\em Gravitation and Cosmology:
Principles and Applications of the General Theory of Relativity}, John Wiley and
sons, (1972).
\bibitem{dkmv} N. Dadhich, S. Kar, S. Mukherjee and M. Visser, 
Phys. Rev. {\bf D 65}, 064004 (2002).
\bibitem{skslssg} S. Kar, S. Lahiri and S. SenGupta, Phys. Rev. {\bf D88}, 123509 (2013),  Erratum {\sf ibid.} Phys. Rev. D 91, 089901 (2015).
\bibitem{grg} S. Kar, S. Lahiri and S. SenGupta, 
Gen. Relativ. Grav. {\bf 47}, 70 (2015).
\bibitem{bk} K. Bronnikov and S-W. Kim, Phys. Rev. {\bf D 67}, 064027 (2003);
F. S. N. Lobo, Physical Review {\bf D 75} 064027 (2007), 
Y. Tomikawa, T. Shiromizu and K. Izumi, Phys. Rev. {\bf D 90}, 126001 (2014);
K. C. Wong, T. Harko and K. S. Cheng, Class. Qtm. Grav. {\bf 14}, 145023 (2011);
F. Parsaei and N. Riazi, Phys. Rev. {\bf D 91}, 024015 (2015).
\bibitem{sms} T. Shiromizu, K. Maeda and M. Sasaki, Phys. Rev. {\bf D62}, 024012 (2000).
\bibitem{lr} L. Randall, Science {\bf 296}, 1422 (2002).

\end{thebibliography}
\end{document}